# TaDaa: real time Ticket Assignment Deep learning Auto Advisor for customer support, help desk, and issue ticketing systems

Leon Feng
Global Business Intelligence
Apple
Cupertino, US
yfeng4@apple.com

Jnana Senapati
Global Business Intelligence
Apple
Cupertino, US
jsenapati@apple.com

Bill Liu*
Global Business Intelligence
Apple
Cupertino, US
bill.liu@apple.com

*Abstract*— **This paper proposes TaDaa: Ticket Assignment Deep learning Auto Advisor, which leverages the latest Transformers models and machine learning techniques quickly assign issues within an organization, like customer support, help desk and alike issue ticketing systems. The project provides functionality to 1) assign an issue to the correct group, 2) assign an issue to the best resolver, and 3) provide the most relevant previously solved tickets to resolvers. We leverage one ticketing system sample dataset, with over 3k+ groups and over 10k+ resolvers to obtain a 95.2% top 3 accuracy on group suggestions and a 79.0% top 5 accuracy on resolver suggestions. We hope this research will greatly improve average issue resolution time on customer support, help desk, and issue ticketing systems.**

*Keywords— Deep Learning, NLP, Transformers, machine learning, help desk, customer service, issue tracking system*

## I. Introduction

Customer support, help desk, and issue tracking ticketing systems are common in most companies/organizations and can benefit from a machine learning solution to improve resolution time and accuracy. We took one issue ticketing systems that receives tens of thousands of new tickets per day with over thousands of assign groups and tenth of thousands of resolvers. We use Machine Learning (ML) especially Deep Learning (DL) techniques to auto-suggest which group to assign, whom to assign, and historically similar tickets for efficient ticket triage and quick resolution.

Generally, there are several issues that reduce ticket resolution efficiency and accuracy. When a new ticket is created, it takes knowledge to find the right team to assign from thousands of teams. The team leader needs to read dozens of tickets per day to assign to right resolver. Resolvers need time to comprehend the issue and find a reference of similar issues and solutions. Any wrong assignments will lead to time wasted on rerouting and reassignment on both Group and Resolver sides.

Most of the previous research focuses on category/group level suggestions, finding similar tickets and providing auto resolution on common issues, by utilizing machine learning and data mining techniques [1-7]. Some researchers also focused on building chatbot for quick response on common issues [8-10]. An AutoML based approach is also investigated [11]. Piero et al. used RNN based Encoder-Combiner-Decoder architecture to auto predict tickets category [1]. DeLucia et al. suggested category, similar tickets, and automatic ticket reply [4]. Powell etc. focus on which organization to assign and automatic ticket reply[12]. Mani etc. built a Deep Learning based question answering system to answer common questions[2]. Gupta et al. used Siamese LSTM to retrieve similar tickets[13]. Some research utilizes user related information to get better suggestions [1, 14, 15]. Some papers investigate on finding similar documents and issues in law cases and biomedical fields [16-19]. Recently, Transformers based models are dominant in most NLP fields [20-26] and are starting to show great value in major computer vision fields [27-29]. Some research has started to utilize Transformers based models in issue ticketing system [30-33].

There is little research investigating Resolver level suggestions. This paper utilizes machine learning techniques to build a slim and low maintenance solution with a focus on providing highly accurate resolver level suggestions. The purpose of this research is not to replace valuable human resolvers, but to help the system reducing issue resolution time and improve resolution quality. This project creates three artifacts that an organization could leverage to provide efficiency, each of which can be used independent of the others:

*1) A Transformer based model that predicts the group a ticket should be assigned to.*

*2) An ensemble of models to predict the resolvers a ticket should be assigned to.*

*3) A Transformer embedding based approximate nearest neighbor's index for fast lookup of similar issues.*

## II. Ticketing system bottleneck summary

Customer support, help desk, and issue tracking ticketing systems are common in most companies/organizations and can benefit from a machine learning solution to improve resolution time and accuracy. One of the issue ticketing systems, receives tens of thousands of new tickets per day with over thousands of assign groups and tens of thousands of resolvers. We use Machine Learning (ML) especially Deep Learning (DL) techniques to auto-suggest which group to assign, whom to assign, and historically similar tickets for efficient ticket triage and quick resolution.

We define three roles in a typical resolution pipeline, each of these roles has a bottleneck in the process associated with it:

- **Reporter** – The individual who has a problem and writes the text describing their issue. **Bottleneck 1**: When a new ticket is created, it takes knowledge to find the right team/group to assign from hundreds/thousands of teams.

- **Group Leader** – A person whose responsibility is to find the right team/person to resolve a ticket. This could be a person doing triage and finding the right group, or once a ticket finds the right group, a manager that gives issues to individual Resolvers. **Bottleneck**

**2**: The Router needs to read dozens/hundreds of tickets per day to assign to right resolver. Bypassing this step entirely from the first-round assignment is the goal.

- **Resolver** – A person with the knowledge and ability to complete the Reporter's issue. **Bottleneck 3**: The Resolver needs time to comprehensively examine the issue and find references to the similar issues and solutions, or even get help from other resolvers.

To fully understand the problem and our solution in an issue ticketing system, the general process of a ticketing system is described below, as shown in Figure 1:

*1) Reporter creates a ticket.*

*2) The ticket must find its way to a group (**Bottleneck 1**)*. This can happen in two ways depending on the system design:

   *a) The user-interface for the Reporter has a mechanism to find the correct group via a menu system. This gets very complex as the number of groups increases.*

   *b) The ticket goes into a global queue and help-desk workers assign them to groups.*

*3) Router of the group, or group leader, reads this ticket then manages it oneself or assigns it to one of the Resolvers in the group (**Bottleneck 2**).*

*4) Resolver reads this ticket and uses their domain knowledge to solve the problem. Potentially, they will check historical solutions and try to solve the ticket (Bottleneck 3); if they are not the right resolver, they will send the ticket back to the queue, potentially to another group.*

*5) Resolver marks the ticket as complete and notifies ticket reporter.*

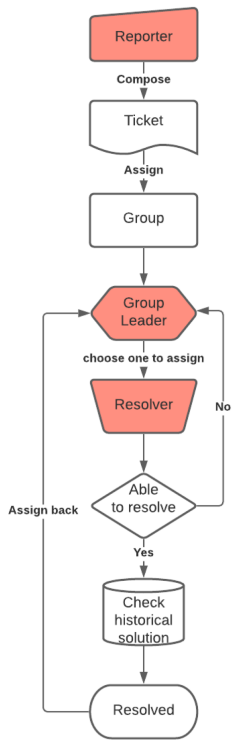

Figure 1: illustration of general ticketing system process

## III. TECHNIQUES

### A. Transformer based classifiers

Transformers are the best performing neural networks in natural language processing since their introduction in 2017 [20]. The dominant paradigm in industry is to utilize Foundation Models [34] like BERT and GPT-3 that are trained on broad data at scale and then adapted to a wide range of downstream tasks. This process, called transfer learning, takes the pre-trained model as a starting point to train a specific task on a specialized dataset. The flow of the model is as follows::

*1) Map sub-words into tokens.*

*2) Map tokens into embeddings.*

*3) Use an encoder to loop through N layers of transformers to create new embeddings.*

*4) Pass the final hidden states to a neural network used for the specific dataset called the classification head.*

*5) The classification head predicts the probability of each class.*

The transformer block is directly imported from a pre-trained model, and the classification head is randomly initialized. The classification head and the transformer parameters are then trained on the ticket corpus to predict the specific classes in our dataset. As shown in Figure 2.

This research compares two key Transformers models BERT and RoBERTa along with their distilled counterparts like DistilBERT, DistilRoBERTa to find the best model [35-38].

The Group classifier and basic Resolver classifier are trained on "group" label and "resolver" label directly by Transformer model.

### B. Transformer encoding based Approximate Nearest Neighbor

Once a ticket has been assigned to a Resolver (Bottleneck 3) we provide a facility of returning the most similar previously resolved tickets for Resolvers to inspect. To do this, each ticket is encoded by Sentence Transformer models into an embedding space (purple box in Figure 2) [39]. The encodings of all previously seen tickets are stored in an index for fast lookup. We use the python package "annoy" from Spotify [40], but many managed services like GCP's Matching Engine exist to do this at scale. At inference time, an Approximate Nearest Neighbor (ANN) algorithm is used to return the most similar tickets in sub-linear time [41], on the order of hundreds of milliseconds.

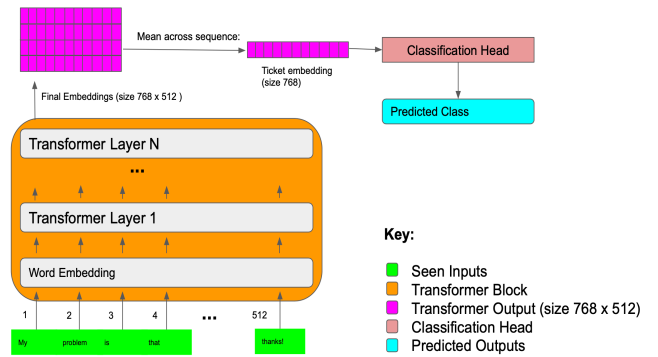

Figure 2: General Transformer architecture

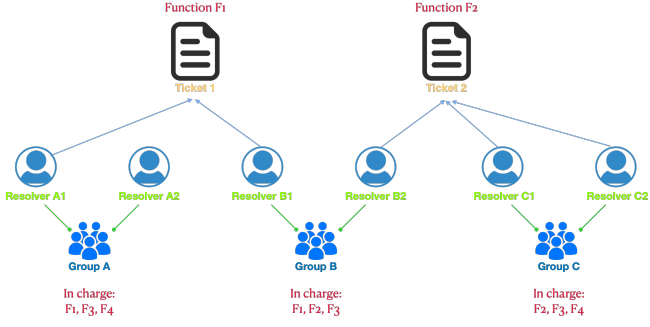

Figure 3: Illustration of ticket one-to-many mapping to groups/resolvers

*C. Top-K accuracy*

Another key idea is that organizations design their groups such that tickets could be resolved by multiple different teams. This duplication is designed on purpose to enhance resolution quality and system functionality, like bridging groups among departments. For other systems, this duplication is due to groups or departments growth causing overlapping functionality. With this duplication issue, the normal accuracy, precision and recall cannot fully represent how good the group-level classifier is. Take the example shown in Figure 3, Ticket 1 can be resolved both by Group A and Group B, the prediction of Group B when the label is Group A will mark this prediction wrong, but prediction of Group B is correct. The same logic holds for Resolvers. Therefore, we use top-K accuracy to evaluate Group and Resolver classifiers.

*D. Ensemble model for Resolver suggestion*

We use four models in an ensemble to predict the Resolver of a ticket. The first model, directly training a Transformer on resolver label is the best but adding in other models helps reduce over-fitting in a single model. The final Resolver suggestions is calculated with predictions from four models below by weighted average ensemble

$$P(R_i) = \sum_i P(R_i|M_j) * W_j \quad (1)$$

where $R_i$ stands of Resolver $i$, $M_j$ sands for Model $j$, $W_j$ stands for weight of Model $j$.

The weights of four models are obtained from training a linear classifier that chooses weights to maximize the accuracy of the combined model.

*1) Suggestions by Resolver classifier*

We use four models in an ensemble to predict the Resolver of a ticket. The first model, directly training a Transformer on resolver label is the best but adding in other models helps reduce over-fitting in a single model. The final Resolver suggestions is calculated with predictions from four models below by weighted average ensemble.

The first model in the ensemble is the Resolver classifier trained by "resolver" label directly. It is also the base model to help evaluate whole model performance.

*2) Suggestions from Resolver-List classifier*

From another perspective, since multiple resolvers can resolve one ticket, we can use similar tickets to find a small list of resolvers as one Resolver-List, that resolved one type of issue or function. We then use the discovered Resolver-List to train an Resolver-List classifier. As Figure 3 shows, Ticket 1 and 2 represents two functions F1 and F2, the possible resolvers Resolver A1, A2 and B1 forms the possible Resolver-List of function F1.

With Resolver-List prediction, the Resolver suggestions can be calculated by

$$P(R_j) = \sum_{k=0}^{n} P(R_j|L_k) * P(L_k) \quad (2)$$

where $R_j$ stands for resolver $j$, $L_k$ stands for Resolver-List $k$, n is the total number of Resolver-List

Since there is no label for Resolver-list, density-based clustering algorithm HDBCAN [42, 43] is used to clustering similar tickets, hence finding Resolver-list automatically. The reasons to choose HDBCAN are:

*a) HDBSCAN does not need to specify number of clusters.*

*b) HDBSCAN will leave data points far from clusters as outlier that will not be forced merging into other clusters.*

Since there could be millions of historical tickets, directly clustering and finding similar tickets would be very time consuming. To achieve faster Resolver-List discovery and real time similar tickets suggestions, Latent Dirichlet Allocation (LDA) is used to clustering all historical tickets into hundreds or thousands of topics [44, 45]. The computational complexity of HDBCAN is O(nlogn), if breaks whole data into hundreds or thousands of topics, theoretically this can greatly reduce clustering computation time.

*3) Suggestions from Group classifier*

A simpler model can be proposed by simply assigning a ticket to the Resolver who handles the most tickets in a group, we can calculate this from the historical data via this formula:

$$P(R_i) = \sum P(R_i|G_j) * P(G_j) \quad (3)$$

where $A_i$ stands for Resolver $i$, and $G_j$ stands for Group $j$.

*4) Suggestions from similar tickets*

Another model that can be added to the ensemble is based on similar tickets. The general idea is:

*a) Rescale the distances of most similar tickets to the interval min and max scale parameters*

$$d_{i,scaled} = \frac{d_i - \theta_{min}}{\theta_{max} - \theta_{min}} \quad (4)$$

where $d_{i,scaled}$ stands for scaled distance of ticket $i$ to the target ticket, $d_i$ stands for original distance of ticket $i$ to the target ticket, $\theta_{min}$ and $\theta_{max}$ stand for the min/max scale parameter.

*b) Give the lowest distance ticket for each Resolver a weight of one. Subsequent tickets from the same Resolver get a weight of*

$$\alpha_{ij} = \begin{cases} \beta^r, & \text{ticket } i \text{ solved by resolver } j \\ 0, & \text{ticket } i \text{ not solved by resolver } j \end{cases} \quad (5)$$

where $\alpha_{ij}$ stands for the weight of ticket $i$ solved by resolver $j$, $\beta$ is the rank discount parameter, $r$ is the rank number of ticket $i$ by resolver $j$.

*c) The score of resolver j is then the product of the scaled distance and the rank weight summed over every ticket that resolver j completed in the similar tickets*

$$s_j = \sum_{i=0}^{n} \frac{\alpha_{ij}}{d_{i,scaled}} \qquad (6)$$

where $s_j$ stands for the score of Resolver $j$.

*d) The probability of resolver j is calculated as the softmax over all the Resolver scores.*

The three parameters ($\theta_{max}$, $\theta_{min}$ and $\beta$) are estimated on the dataset to minimize the cross entropy log-loss of the model. The two scaling parameters $\theta_{max}$ and $\theta_{min}$ control the spread of inverted distances to allow the softmax to give more probability to closer tickets. The rank discount parameter $\beta$ controls the amount of score to give to resolvers that solved multiple similar tickets. The total increase in score for a resolver solved many tickets is bounded by $1 + 1/\beta$. with geometric sum of sequence

$$\omega = 1 + \sum_{i=1}^{n} \frac{1}{r^i} \qquad (7)$$

where $\omega$ is the total increase in score.

## IV. MODEL PIPELINE ARCHITECTURE

There are Training pipeline and inference pipeline to separate functionality in Production.

### A. Training pipeline

The training pipeline process is shown in Figure 4. After basic data cleaning, historical tickets are used to directly train the Group classifier and the Resolver classifier. The tickets are also passed to the LDA model, HDBSCAN, Resolver-List classifier pipeline. Finally, transformer-based embedded historical tickets are used to generate ANN index for fast lookup.

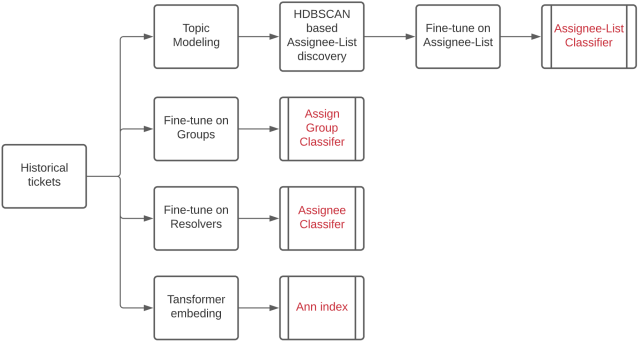

Figure 4. General training pipeline

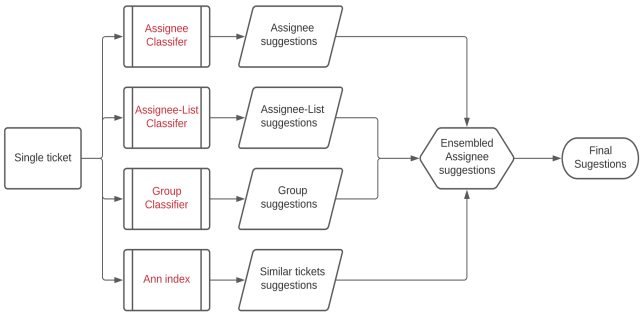

Figure 5. General inference pipeline

### B. Inference pipeline

The inference pipeline is shown in Figure 5. After basic data cleaning, this new single ticket is sent to four sub-pipelines: Resolver Pipeline, Group pipeline, Resolver-List pipeline and similar tickets pipeline. The group-level prediction come solely from the Group classifier. The similar tickets pipeline encodes the ticket with a transformer and then passes the encoding to an Approximate Nearest Neighbor index that returns the most similar tickets. The resolver prediction is an ensemble of the Resolver pipeline, Group pipeline, Resolver-List pipeline and similar tickets pipeline.

## V. DATASET AND EXPERIMENTAL ENVIRONMENT

The dataset is a sample of 20 days of tickets from one ticketing system with 203300 tickets. The entire machine learning system only uses three columns: "group", "resolver", and "description". Where "group" is the team from which the ticket is resolved by, "resolver" is the specific person who solved the ticket, and "description" is the text of the ticket. After removing tickets with non-sense "description" and empty value on "resolver"/"group", 144600 tickets are left to be split into train, validation, and test with 8:1:1 ratio. The experiments were performed on an cluster with GPUs. All experiments are running on same settings of 10 logical CPU cores, 100GB Ram and dual V100 GPUs.

## VI. EXPERIMENTS AND RESULTS

### A. Group suggestions

The Top-K accuracy of the Group classifier run with multiple Transformer architectures is shown in Table I. It shows RoBERTa Transfomer get highest Top-K accuracy with big training time increase and slight inference time increase. With RoBERTa model, if we choose the most likely group, the 1st choice prediction accuracy is around 82%. Due to the overlap in functionality between groups we think a top-3 accuracy of 95.5% is a better metric for the success of the system. The increase of about 12.8% between Top-1 and Top-3 shows how severe the ambiguous meaning in group labels. It also demonstrates Top-3 suggestions from Transformers-based Group classifier is quite reliable.

A typical workflow is to have the user select which group to send their issue to. When the number of potential groups becomes large, for example hundreds or thousands, presenting this to a user becomes daunting. The user interface for a machine learning powered ticket assignment becomes more streamlined, the user types the request and when finished the UI presents the top-k groups along with the description of what type of problems the group handles.

### B. Ensembled Resolver suggestions

In the sample dataset, the average resolvers per group is around seven. Therefore, we select the Top-5 accuracy to evaluate the performance of Resolver suggestions. The calculated Top-K accuracy of Resolver suggestions is shown in Table 2, each of the models uses the RoBERTa transformer.

Using our preferred metric of Top-5 accuracy, Model 1 got an accuracy of 77.1%, Model 2 got an accuracy of 71.9%, Model 3 got an accuracy of 77.2%. and Model 4 got an accuracy of 40.4%. The ensemble improved on the best model by 1.9% with an accuracy of 79.0%.

There is an important difference between the Top-1 accuracy of Model 1 and Model 3. Model 3 simply predicts the group and then predicts whoever has resolved the most tickets in the group. Model 1 on the other hand associates the context in each ticket with individual resolvers and predicts the highest probability person. The improvement of 19.6% to 40.4% reflects that there is very substantial specialization of tasks within each group that our transformer is picking up on.

*C. Training time and inference time*

The average running time on whole pipeline training with early-stopping is around 37.5 hours on RoBERTa transformer-based solution on duel-V100 GPU. This shows the whole pipeline could not be trained on daily basis but two-day bases with single or dual V100 GPU(s). However, with four V100 GPUs, the whole pipeline could be trained on daily basis to refresh new groups and new resolvers.

If allowing slight sacrifice of Top-K accuracy, the whole pipeline can be trained on daily basis with single V100 level GPU by simply replacing RoBERTa model to distilRoBERTa model.

The average inference time for each model and a call to the Approximate Nearest Neighbor index for 500 similar tickets is only 0.135 second on 6 cores 2.6GHz MacBook Pro CPU and 0.05 seconds on a GPU server. The typical inference time is only 0.330 second on 6 cores 2.6GHz MacBook Pro CPU. The method is well suited to a REST API for real time auto suggestions.

TABLE I.   TOP 3 ACCURACY OF GROUP CLASSIFIER

| Model Name | Top 1 | Top 2 | Top 3 | Training Time (h) GPU | Inference Time (s) CPU |
|---|---|---|---|---|---|
| BERT | 0.801 | 0.915 | 0.951 | 8.0 | 0.065 |
| distilBERT | 0.803 | 0.923 | 0.95 | 4.5 | 0.045 |
| **RoBERTa** | **0.824** | **0.925** | **0.952** | 12.5 | **0.065** |
| distilRoBERTa | 0.808 | 0.92 | 0.951 | 4.5 | 0.045 |

TABLE II.   TOP-K ACCURACY OF RESOLVER MODEL

| Model | Top 1 | Top 3 | Top 5 | Training Time (h) GPU | Inference Time (s) CPU |
|---|---|---|---|---|---|
| 1. Resolver Classifier | 0.494 | 0.690 | 0.771 | 12.5 | 0.065 |
| 2. Resolver-List Classifier | 0.495 | 0.689 | 0.772 | 12.5 | 0.065 |
| 3. Group Classifier | 0.196 | 0.309 | 0.404 | 12.5 | 0.065 |
| 4. Similar tickets Classifier | 0.436 | 0.638 | 0.719 | NA | 0.135 |
| Ensemble (4 models) | **0.510** | **0.705** | **0.790** | 37.5 | **0.33** |

## VII. CONCLUSION

This paper proposes TaDaa, a slim solution that focusing on improving overall suggestions, especially on Resolver level suggestions for customer support, help desk and issue ticketing systems. We demonstrated a common issue that one ticket can be resolved by multiple resolvers from multiple groups. TaDaa can significantly improve suggestion accuracy and reduce ticket resolution time for the entire system. TaDaa is also a general solution in which any of the subcomponents can be utilized it to improve accuracy and efficiency.